\pdfoutput=1
\documentclass[11pt]{article}

\usepackage[utf8]{inputenc}
\usepackage[T1]{fontenc}
\usepackage{lmodern}

\usepackage[margin=1in]{geometry}
\usepackage{microtype}

\usepackage{amsmath,amssymb,mathtools}
\usepackage{booktabs}
\usepackage{multirow}
\usepackage{siunitx}
\usepackage{enumitem}
\usepackage{graphicx}
\usepackage{tikz}
\usetikzlibrary{positioning,arrows.meta,matrix,shapes.geometric,calc}

\usepackage{algorithm}
\usepackage{algpseudocode}

\usepackage{xcolor}
\usepackage{url}
\usepackage[authoryear,round]{natbib}
\usepackage[hidelinks]{hyperref}

\providecommand{\Description}[1]{}

\title{Collective Recourse for Generative Urban Visualizations}

\author{Rashid Mushkani\\
Université de Montréal\\
Mila -- Quebec AI Institute}

\date{} 

\begin{document}
\maketitle

\begin{abstract}
Text-to-image diffusion models help visualize urban futures but can amplify group-level harms. We propose \emph{collective recourse}: structured community ``visual bug reports'' that trigger fixes to models and planning workflows. We (1) formalize collective recourse and a practical pipeline (report, triage, fix, verify, closure); (2) situate four recourse primitives---counter-prompts, negative prompts, dataset edits, reward-model tweaks---within the diffusion stack; (3) define mandate thresholds via a \emph{mandate score} combining severity, volume saturation, representativeness, and evidence; and (4) evaluate a synthetic program of 240 reports. Prompt-level fixes were fastest (median $2.1$--$3.4$ days) but less durable (21--38\% recurrence); dataset edits and reward tweaks were slower (13.5 and 21.9 days) yet more durable (12--18\% recurrence) with higher planner uptake (30--36\%). A threshold of $0.12$ yielded 93\% precision and 75\% recall; increasing representativeness raised recall to 81\% with little precision loss. We discuss integration with participatory governance, risks (e.g., overfitting to vocal groups), and safeguards (dashboards, rotating juries).
\end{abstract}


\section{Introduction}
Text-to-image diffusion models \citep{Ho2020DDPM,Rombach2022LDM} are entering planning practice to render streetscape alternatives, infill options, and design guidelines. In this study, we focus on Stable Diffusion XL (SDXL) because it is open source; however, other image generators are also used in this domain. Emerging work in urban informatics and design indicates both promise and fragility \citep{Jiang2024GenUrbanDesignReview,Xu2024GenAITwins}. Beyond aesthetic plausibility, audits of image generators report systematic social biases: demographic stereotyping and omission \citep{Luccioni2023StableBias}, amplified occupational gender biases \citep{Wang2024NewJobNewGender}, prompt-surface vulnerabilities \citep{Struppek2023Homoglyphs}, and, in large-scale evaluations, toxic or racialized outputs \citep{Schneider2025StableDiffusionToxicity}. These issues are often linked to training data composition and uneven coverage of everyday civic life \citep{Birhane2021MultimodalDatasets}. In civic contexts, such distortions matter because renderings can influence deliberation when used as decision inputs.

Algorithmic \emph{recourse} offers those affected by a model pathways to change outcomes, originally for classification \citep{Ustun2019ActionableRecourse,Karimi2022SurveyRecourse}. Counterfactual explanations propose minimally altered alternatives \citep{Wachter2017Counterfactual}, with later methods emphasizing diversity and feasibility \citep{Mothilal2020DiCE,Poyiadzi2020FACE}. Importantly, recourse fairness can diverge from prediction fairness \citep{VonKugelgen2022FairRecourse}. In generative urban imagery, harms are typically \emph{group-level} and recurring across prompts and contexts, suggesting a shift from individual to \emph{collective} recourse tied to model lifecycle and municipal governance \citep{Mushkani2025AIES_CoProducingAI}.

We propose \textbf{collective recourse} for generative urban visualizations: organized community inputs become structured \emph{visual bug reports} that, when meeting clear thresholds, trigger specific fixes in the model stack and propagate into planning artifacts. Our design draws on traditions of participatory planning \citep{Arnstein1969Ladder,Healey1997CollaborativePlanning,Sieber2006PPGIS}, taxonomies of participatory power \citep{Fung2006VarietiesParticipation}, and empirical studies of user-driven audits \citep{Li2023UserAudits,DeVrio2024PowerFromBelow,Mushkani2025ICML_RightToAI}. Rather than treating transparency as an end in itself \citep{Ananny2018SeeingWithoutKnowing}, we specify a pipeline with service-level targets, subgroup verification, and independent citizen review.

\paragraph{Research questions.}
We study two questions that bridge technical remedies and institutional mechanisms. \textbf{RQ1}: Which \emph{recourse primitives}---counter-prompts, negative prompts, dataset edits, or reward-model tweaks---most quickly reduce \emph{group-level visual harms} while limiting recurrence? This extends counterfactual insights \citep{Wachter2017Counterfactual,Mothilal2020DiCE,Poyiadzi2020FACE} to generative settings and engages recent work on aligning diffusion models with preferences \citep{Christiano2017RLHF,Wallace2024DiffusionDPO}. \textbf{RQ2}: What \emph{collective thresholds} constitute a \emph{mandate} for mitigation, and how do representativeness and evidence quality trade off with speed and precision, given equity concerns in access to recourse \citep{VonKugelgen2022FairRecourse} and institutionalizing participatory authority \citep{Fung2006VarietiesParticipation}?

\paragraph{Contributions.}
We: (1) formalize collective recourse for generative imagery and specify a practical pipeline from reporting to triage, fix selection, verification, and closure; (2) introduce a typology of four recourse primitives in the diffusion stack and discuss their mechanisms and limits; (3) define mandate thresholds that translate collective signals into enforceable actions with transparent precision--recall trade-offs; and (4) conduct a simulation-based evaluation on a synthetic program of 240 community reports, quantifying mitigation latency, recurrence, subgroup satisfaction, and downstream policy uptake, and comparing our approach to red-teaming and third-party audits \citep{Ganguli2022RedTeamLMs,Longpre2023SafeHarbor,Raji2022OutsiderOversight,Raji2020ClosingGap}.

\section{Background and related work}
\subsection{Generative models in planning and their biases}
Diffusion models \citep{Ho2020DDPM,Rombach2022LDM} are now common tools for depicting alternatives. Surveys document their rise in urban design and digital twins \citep{Jiang2024GenUrbanDesignReview,Xu2024GenAITwins}, including street-level and satellite synthesis. Audits report persistent social biases: under-representation and stereotyping, skewed cultural markers, and occupational gender bias \citep{Luccioni2023StableBias,Wang2024NewJobNewGender}. Prompt-surface vulnerabilities can elicit hidden behaviors \citep{Struppek2023Homoglyphs}, and evaluations note toxicity and racialized portrayals in some outputs \citep{Schneider2025StableDiffusionToxicity}. Training data analyses implicate multimodal corpora with problematic content and uneven coverage \citep{Birhane2021MultimodalDatasets}. Because civic renderings shape perceptions of safety, desirability, and legitimacy, these risks are salient for planning, where perception datasets such as Place Pulse have long probed visual judgments \citep{Naik2014Streetscore,Salesses2013CollaborativeImage,Mushkani2025JUM_MontrealStreets,MushkaniKoseki2025Habitat_StreetReview}.

\subsection{From individual recourse to collective recourse}
Recourse research originally addressed individuals facing adverse automated decisions \citep{Ustun2019ActionableRecourse,Karimi2022SurveyRecourse}. Counterfactual explanations communicate ``what to change'' \citep{Wachter2017Counterfactual}, with later work prioritizing diverse and feasible options \citep{Mothilal2020DiCE,Poyiadzi2020FACE}. A purely individual framing does not capture \emph{recurring, group-level} harms in generative systems, and equitable access to remedies is itself a fairness concern \citep{VonKugelgen2022FairRecourse}. We extend this line by defining collective recourse: groups file structured visual bug reports; institutions adopt thresholds that \emph{trigger} mitigations; and changes are verified on subgroup test sets and reviewed by citizen panels.

\subsection{Documentation, governance, and the limits of transparency}
Model Cards and Data Statements document scope and limitations \citep{Mitchell2019ModelCards,Bender2018DataStatements}; Datasheets formalize dataset documentation \citep{Gebru2021Datasheets}. Our pipeline operationalizes these artifacts: dataset edits update datasheets; model-level changes update model cards; closures are logged publicly. We also adopt an end-to-end audit mindset \citep{Raji2020ClosingGap}: structured intake, triage, verification, and independent review. Empirical work suggests that participatory audits can surface issues that experts miss and can precipitate institutional change \citep{Li2023UserAudits,Howell2024ReflectiveDesign,DeVrio2024PowerFromBelow}. Our rotating citizen juries and mandate thresholds are informed by these lessons and by democratic design frameworks that specify \emph{who} participates, \emph{how}, and with \emph{what authority} \citep{Fung2006VarietiesParticipation}.

\subsection{Red-teaming and third-party oversight}
Red-teaming formalizes adversarial testing for generative models \citep{Ganguli2022RedTeamLMs}. Assessments identify blind spots (e.g., limited context coverage and weak ties to action). Safe-harbor provisions aim to broaden participation and protect testers \citep{Longpre2023SafeHarbor}. Outsider oversight and third-party audits help provide structural accountability \citep{Raji2022OutsiderOversight}. We position collective recourse as complementary: community members act as a distributed red team through binding bug reports with clear escalation routes and institutional links.

\section{Problem formulation}
Let a \emph{visual bug report} $b$ reflect a recurring harm experienced by a \emph{group} $g$ when generating images for an urban \emph{context} $c$ (e.g., a borough, typology, or program), with structured metadata
\begin{equation*}
b = \langle g, c, \text{prompt(s)}, \text{evidence}, \text{harm type}, s, n, r, q \rangle,
\end{equation*}
where $s\in[0,1]$ is severity, $n$ the count of unique reporters, $r\in[0,1]$ a \emph{representativeness index} (e.g., stratified alignment with population marginals), and $q\in[0,1]$ evidence quality. This structure makes explicit fairness-sensitive dimensions of recourse \citep{VonKugelgen2022FairRecourse} and aligns with participatory governance principles that link authority to participation \citep{Fung2006VarietiesParticipation}.

We define a \emph{mandate score}
\begin{equation}
M(b) \;=\; s \cdot \underbrace{\bigl(1-\exp(-n/\tau_n)\bigr)}_{\text{volume saturation}} \cdot r \cdot q,
\label{eq:mandate}
\end{equation}
with $\tau_n$ controlling saturation. We consider a mitigation \emph{mandated} when $M(b)\ge \tau_M(h)$ for harm type $h$ (e.g., stereotyping vs.\ omission), with $\tau_M$ chosen to meet target precision/recall trade-offs (Section~\ref{sec:eval}). Because $r$ captures representativeness rather than volume, the design aims to reduce over-weighting vocal but unrepresentative coalitions and invites procedural innovations (e.g., rotating juries) that can increase $r$ without inflating $n$.

\subsection{Recourse primitives in the diffusion stack}
We consider four primitives that operate at distinct layers and exhibit different speed--durability profiles. \textbf{Counter-prompts} add protective, context-aware language (e.g., ``include universal design features; retain vernacular signage''), offering low-latency responses aligned with counterfactual reasoning \citep{Wachter2017Counterfactual}. \textbf{Negative prompts} exclude unwanted motifs; their mechanics and limits are increasingly characterized \citep{Ban2024NegativePrompts}. \textbf{Dataset edits} curate and augment underrepresented scenes, objects, or communities; they are documented via datasheets and generally produce more durable changes \citep{Bender2018DataStatements,Gebru2021Datasheets,Mushkani2025ICML_LIVS}. \textbf{Reward-model tweaks} update preference or reward models to steer generation at sampling time, analogous to RLHF for language \citep{Christiano2017RLHF}; direct preference optimization for diffusion models offers promising alignment leverage \citep{Wallace2024DiffusionDPO}. Because prompt surfaces can be exploited \citep{Struppek2023Homoglyphs}, we treat prompt-level fixes as necessary first responses rather than complete solutions; structural fixes are prioritized when harms recur.

\subsection{Verification on subgroup test sets}
For each harm-type $\times$ context, we maintain a \emph{subgroup visual eval set} (SVES) of templated prompts and reference imagery, seeded by urban-perception cues \citep{mushkani2025streetreviewparticipatoryaibased}. Verification re-generates the SVES pre- and post-mitigation and tests recurrence against harm-specific tolerances. Because community participation is central to surfacing harms \citep{Li2023UserAudits,Howell2024ReflectiveDesign}, we encourage \emph{safe harbor} reporting \citep{Longpre2023SafeHarbor} and publish aggregate dashboards with stratified breakdowns while suppressing small cells.

\section{Pipeline design}
\begin{figure}[t]
\centering
\resizebox{\textwidth}{!}{%
\begin{tikzpicture}[node distance=7mm and 9mm, font=\small]
\tikzstyle{block}=[draw, rounded corners, align=center,
                   minimum height=8mm, inner sep=3pt, fill=gray!10]

\node[block] (report) {Community submission \\ \textbf{Visual bug report}};
\node[block, right=of report] (triage) {Triage \& routing \\ (harm type, context $c$)};
\node[block, right=of triage] (fix) {Select fix \\ CP / NP / DE / RT};
\node[block, right=of fix] (verify) {Verify on SVES \\ recurrence $\downarrow$?};
\node[block, right=of verify] (review) {Citizen review \\ (rotating jury)};
\node[block, right=of review] (close) {Closure \& \\ public dashboard};

\draw[-{Latex}] (report) -- (triage);
\draw[-{Latex}] (triage) -- (fix);
\draw[-{Latex}] (fix) -- (verify);
\draw[-{Latex}] (verify) -- node[above]{pass/fail} (review);
\draw[-{Latex}] (review) -- (close);

\node[block, below=of fix] (policy) {Planning link: \\ design guides, RFQs, \\ PB criteria};
\draw[-{Latex}] (verify) -- (policy);
\draw[-{Latex}] (policy) -- (close);
\end{tikzpicture}%
}
\caption{Collective recourse pipeline tying community reports to model fixes and planning workflows. Citizen review and public dashboards operationalize accountability \citep{Arnstein1969Ladder,Raji2020ClosingGap}.}
\Description{Flowchart with six main steps laid out left to right: (1) Community submission: visual bug report; (2) Triage and routing by harm type and context $c$; (3) Select fix (CP/NP/DE/RT); (4) Verify on SVES with the question “recurrence ↓?”; (5) Citizen review by a rotating jury (arrow labeled pass/fail); (6) Closure and public dashboard. A secondary box below “Select fix,” labeled “Planning link: design guides, RFQs, PB criteria,” receives an arrow from “Verify on SVES” and sends an arrow to “Closure \& public dashboard.” The figure illustrates how community reports propagate through technical validation and public oversight to closure and transparency.}
\label{fig:pipeline}
\end{figure}
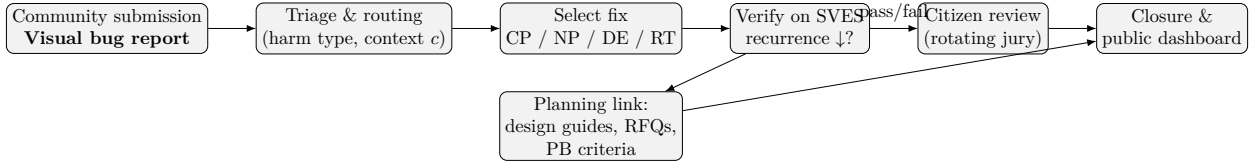

\paragraph{Coordinated advocacy channel with enforcement.}
A dedicated intake and triage team (city, vendor, and community liaisons) publishes service-level targets by severity. Visual bug reports are clustered by harm type and context, then routed to candidate fixes. Mandate thresholds (Eq.~\ref{eq:mandate}) transform collective signals into commitments. Each closure updates public-facing documentation: model cards for model-level changes; datasheets for data-level changes \citep{Mitchell2019ModelCards,Gebru2021Datasheets}. We integrate these artifacts with planning workflows by linking verified fixes to design guides, procurement language, and participatory budgeting criteria, aligning governance with model improvement rather than isolating it as a separate audit silo \citep{Raji2020ClosingGap}.

\paragraph{Citizen review and division of labor.}
We convene rotating mini-publics (citizen juries) to review closures, adjudicate disputed cases, and audit representativeness. Evidence suggests that informal, qualitative audits by non-experts can surface issues that experts miss \citep{Howell2024ReflectiveDesign} and that user-driven investigations can catalyze organizational change \citep{Li2023UserAudits,DeVrio2024PowerFromBelow}. The jury model clarifies roles: communities \emph{protect} by flagging harms; auditors and vendors \emph{strengthen} by verifying fixes and updating documentation; administrators \emph{pressure} by enforcing mandates—consistent with taxonomies of participatory power \citep{mushkani2025negotiativealignmentembracingdisagreement}.

\section{Evaluation: simulation study}\label{sec:eval}
\subsection{Setup}
We simulate a year-one program receiving $N{=}240$ reports across six subgroups (Women, Seniors, LGBTQ+, Newcomers, African diaspora, Neighborhood organizations).\footnote{This section uses synthetic data to illustrate choice-of-fix and threshold trade-offs; no human subjects data were collected. Code and random seeds can be shared in a reproduction package.} For each report we sample severity $s$, reporters $n$, representativeness $r$, and evidence quality $q$. Triage assigns a candidate fix type (CP/NP/DE/RT) based on severity bands and early signals of prompt steerability. Outcomes are tracked as continuous metrics: \emph{mitigation latency} (days from intake to verified pass on the SVES); \emph{recurrence} (probability that the problem reappears on the SVES within 30 days); \emph{subgroup satisfaction} (a Likert proxy for sign-off by the affected group); and \emph{policy adoption} (whether planners adopt a related textual or visual change within six months). We then study mandate thresholds by comparing $M(b)$ (Eq.~\ref{eq:mandate}) to a hidden simulator label indicating “true need,” enabling precision and recall estimates under alternative thresholds and exploring how increases in representativeness (e.g., via rotating juries) shift performance.

\subsection{Results}
Prompt-level actions (counter-prompts and negative prompts) are consistently the \emph{fastest} responses (median $2.1$ and $3.4$ days, respectively) but exhibit higher recurrence (21--38\%). This aligns with known prompt-surface vulnerabilities \citep{Struppek2023Homoglyphs} and the brittleness of term-based exclusions \citep{Ban2024NegativePrompts}. Dataset edits and reward-model tweaks are slower (median $13.5$ and $21.9$ days) but deliver more durable mitigation (12--18\% recurrence) and higher downstream policy adoption (30--36\%), consistent with the expectation that structural changes generalize better and are easier to encode in planning artifacts \citep{Bender2018DataStatements,Mitchell2019ModelCards}. Table~\ref{tab:fix} summarizes these patterns.

\begin{table}[t]
\centering
\caption{Synthetic outcomes by recourse primitive ($N{=}240$ reports).}
\label{tab:fix}
\begin{tabular}{lrrrrrr}
\toprule
Fix type & $n$ & Median latency & IQR & Recurrence & Mean sat. & Policy adopted \\
\midrule
Counter-prompt & 97 & 2.1 d & 1.3 d & 38.1\% & 3.97 & 10.3\% \\
Negative prompt & 47 & 3.4 d & 1.3 d & 21.3\% & 4.16 & 14.9\% \\
Dataset edit    & 63 & 13.5 d & 6.5 d & 17.5\% & 4.71 & 30.2\% \\
Reward tweak    & 33 & 21.9 d & 8.1 d & 18.2\% & 4.70 & 36.4\% \\
\bottomrule
\end{tabular}
\end{table}

Mandate thresholds provide clear, contestable triggers (Table~\ref{tab:thresholds}). For early-stage programs prioritizing \emph{precision}, $\tau_M{=}0.16$ yields $96.4\%$ precision at $51.9\%$ recall; programs emphasizing \emph{recall} may prefer $\tau_M{=}0.12$ (92.9\% precision, 75.0\% recall). Increasing representativeness by $\Delta r{=}0.05$ through rotating juries raises recall to $81.4\%$ with a minor precision change (from $92.9\%$ to $91.4\%$), suggesting that procedural innovations can improve coverage without substantial loss in precision.

\begin{table}[t]
\centering
\caption{Mandate threshold sensitivity (synthetic).}
\label{tab:thresholds}
\begin{tabular}{rrrr}
\toprule
$\tau_M$ & Flagged (share) & Precision & Recall \\
\midrule
0.08 & 73.8\% & 83.6\% & 94.9\% \\
0.12 & 52.5\% & 92.9\% & 75.0\% \\
0.16 & 35.0\% & 96.4\% & 51.9\% \\
0.20 & 20.4\% & 100.0\% & 31.4\% \\
0.24 & 10.4\% & 100.0\% & 16.0\% \\
0.28 & 3.8\%  & 100.0\% & 5.8\% \\
\bottomrule
\end{tabular}
\end{table}

Subgroup analyses show median latency ranging from $2.8$ to $4.7$ days, with Seniors experiencing the lowest recurrence (13.9\%) and Women slightly higher latency (4.7 days). These differences underscore the value of public dashboards and stratified service levels that adapt resources to close gaps, complementing formal treatments of recourse fairness \citep{VonKugelgen2022FairRecourse}.

\begin{table}[t]
\centering
\caption{Subgroup outcomes (synthetic; select indicators).}
\label{tab:subgroups}
\begin{tabular}{lrrrr}
\toprule
Subgroup & $n$ & Median latency & Recurrence & Policy adopted \\
\midrule
African diaspora  & 36 & 4.0 d & 30.6\% & 19.4\% \\
LGBTQ+            & 37 & 3.0 d & 35.1\% & 16.2\% \\
Neighborhood orgs & 39 & 2.8 d & 30.8\% & 23.1\% \\
Newcomers         & 49 & 3.4 d & 26.5\% & 20.4\% \\
Seniors           & 36 & 4.0 d & 13.9\% & 22.2\% \\
Women             & 43 & 4.7 d & 23.3\% & 18.6\% \\
\bottomrule
\end{tabular}
\end{table}

\subsection{Comparison to red-teaming and third-party audits}
Our pipeline can be viewed as an institutionalized, user-driven red team with explicit closure criteria. Traditional red-teaming is valuable for stress-testing \citep{Ganguli2022RedTeamLMs} but can struggle to connect findings to persistent, context-specific harms and to formal authority for remedies. Safe-harbor provisions \citep{Longpre2023SafeHarbor} and outsider oversight \citep{Raji2022OutsiderOversight} broaden participation and sustain pressure. We embed these ideas into municipal workflows with mandate thresholds, subgroup verification, and public dashboards.

\section{Methods in practice: from bug reports to planning}
\subsection{Report schema, triage, and fix selection}
Each report captures prompts, generated images, context, harm type, a severity rubric, optional reporter demographics, substantiating evidence, and a proposed remedy. Triage proceeds in two passes. First, we detect cross-report patterns by clustering on context and harm type to identify candidates for structural fixes when harms recur. Second, we test prompt steerability on the SVES; if targeted counter- or negative prompts reliably reduce recurrence, we deploy them as first responses. Persistent omissions or misrepresentations traceable to data coverage trigger dataset curation and augmentation documented via datasheets \citep{Gebru2021Datasheets}. If misalignment remains despite adequate coverage, we update reward models, drawing on diffusion-specific alignment techniques such as direct preference optimization \citep{Wallace2024DiffusionDPO} in addition to RLHF-style objectives \citep{Christiano2017RLHF}. Model cards record scope and limitations \citep{Mitchell2019ModelCards}, and every change is visible on a public dashboard.

\subsection{Verification, closure, and planning linkages}
Verification re-runs the SVES for the affected subgroup; closure requires recurrence below a harm-specific tolerance, subgroup sign-off above a satisfaction threshold, and citizen-jury approval. Verified fixes propagate to design guides, RFQ language, and participatory budgeting criteria so that technical corrections are reflected in institutional artifacts. Public dashboards log case history, thresholds, decisions, and documentation updates, operationalizing an end-to-end accountability loop \citep{Raji2020ClosingGap}. We view red-teaming signals and external audits as complementary inputs rather than stand-alone solutions \citep{Longpre2023SafeHarbor,Raji2022OutsiderOversight}.

\begin{algorithm}[t]
\caption{Collective Recourse Triage (sketch)}
\begin{algorithmic}[1]
\Require Report $b$ with $(s,n,r,q)$; SVES; policy linkages
\State Compute $M(b)$ via Eq.~\ref{eq:mandate}; \textbf{if} $M(b)\!<\!\tau_M$ \textbf{then} queue for monitoring
\State Detect pattern across reports (clustering by context, harm type)
\If{SVES shows prompt steerability} \State fix $\gets$ CP/NP
\ElsIf{coverage gap identified (datasheet)} \State fix $\gets$ DE
\Else \State fix $\gets$ RT
\EndIf
\State Implement fix; re-run SVES; log metrics (latency, recurrence)
\State Citizen review; link verified fixes to design guides / RFQs
\end{algorithmic}
\end{algorithm}

\section{Discussion}
\paragraph{Answer to RQ1.}
Prompt-level fixes act as \emph{first responses}: fast, easy to operationalize in workshops, and amenable to templating \citep{mushkani2025wedesigngenerativeaifacilitatedcommunity}. They are also more prone to recurrence and surface-level attacks \citep{Struppek2023Homoglyphs}. Structural fixes—dataset edits and reward-model alignment—are slower but appear more durable and easier to encode in institutional artifacts. Preference-optimization approaches for diffusion models \citep{Wallace2024DiffusionDPO} may help correct subtle omissions that are hard to guardrail lexically.

\paragraph{Answer to RQ2.}
Mandate thresholds translate collective signals into enforceable action and provide clear levers for governance. In our simulation, $\tau_M{=}0.12$ balances urgency (75\% recall) and rigor (93\% precision); $\tau_M{=}0.16$ suits resource-constrained settings prioritizing precision. Modestly increasing representativeness through rotating juries improves recall with limited precision loss, echoing concerns about equitable access to remedies \citep{VonKugelgen2022FairRecourse} and empowered participation \citep{Fung2006VarietiesParticipation}.

\paragraph{Risks, mitigations, and limits.}
Risks include \emph{overfitting to vocal groups} (mitigated by the representativeness index $r$, jury rotation, and stratified recruitment), \emph{gaming and astroturfing} (addressed via rate-limiting, deduplication, and evidence-quality checks; safe harbor policies protect good-faith reporters \citep{Longpre2023SafeHarbor}), and the \emph{transparency trap} where dashboards substitute for action \citep{Ananny2018SeeingWithoutKnowing}. Our pipeline incorporates transparency alongside thresholds, verification, and citizen review. We also separate exploratory ideation from evidentiary renderings to avoid misleading inputs to deliberation.

\paragraph{Integration with municipal workflows.}
Cities can embed collective recourse in existing engagement portals; link verified fixes to design guideline updates and RFQ evaluation criteria; and publish rolling dashboards with subgroup SLAs. In doing so, they align technical governance with collaborative planning traditions \citep{Healey1997CollaborativePlanning} and with calls for structured outsider oversight \citep{Raji2022OutsiderOversight}.

\section{Limitations}
Our evaluation uses a simulation to illustrate design trade-offs; effect sizes will vary by toolchain, data provenance, and civic capacity. While SVES verification draws inspiration from perception datasets \citep{Naik2014Streetscore,Salesses2013CollaborativeImage}, constructing robust subgroup test sets for each harm type and context remains a community-intensive endeavor. Future work will field-test the pipeline with partners; build shared SVES resources seeded with urban-perception cues; compare diffusion-specific preference optimization against data curation at scale \citep{Wallace2024DiffusionDPO}; and study the labor and equity costs of citizen review, building on user-driven audits \citep{Li2023UserAudits,Howell2024ReflectiveDesign,DeVrio2024PowerFromBelow}. We also plan to examine legal and organizational enablers for safe participation \citep{Longpre2023SafeHarbor,mushkani2025urbanaigovernanceembed} and to evaluate dashboards as boundary objects for third-party auditors \citep{Raji2022OutsiderOversight}.

\section{Conclusion}
Collective recourse reframes who gets to correct generative urban visualization systems and \emph{how} those corrections propagate into both models and planning practice. Our pipeline makes harms reportable, fixes auditable, and closures accountable. Simulations suggest a division of labor: prompt-level interventions for speed; data- and reward-level changes for durability and policy impact. Mandate thresholds translate collective signals into enforceable action, especially when combined with rotating citizen juries. Situating this within established participatory planning traditions and contemporary audit practices may help cities realize the benefits of generative tools while reducing group-level harms.

\bibliographystyle{plainnat}
\bibliography{references}

\end{document}